\documentstyle[12pt]{article}
\newcommand{\figurebox}[2]{\mbox{\vbox to#2in{\hbox to #1in{\hfil}
\vfil}}}

\def\phipi{\phi_{\pi}(u)}
\def\varsig{\varphi_{\sigma}(u)}

\def\rhop{\rho_{ \perp}}
\def\sigrh{\sigma_{\xi \nu}}
\def\varep{\varepsilon_{\mu \nu \lambda \sigma}}
\def\frho{F_{ \pi \rho} ( Q^{2})}
\def\lo{\langle 0 |}

\def\gmmu{\gamma _{\mu}}
\def\gmnu{\gamma _{\nu}}
\def\gmf{\gamma _{5}}
\newcommand{\beq}{\begin{equation}}
\newcommand{\eeq}{\end{equation}}
\newcommand{\bea}{\begin{eqnarray}}
\newcommand{\eea}{\end{eqnarray}}

\begin{document}
\renewcommand{\thefootnote}{\fnsymbol{footnote}}
                                        \begin{titlepage}
\begin{flushright}
MPI--PhT/94--8\\
TAUP--2137-94 \\
hep-ph/9402270
\end{flushright}
\vskip0.8cm
\begin{center}
{\LARGE
Soft contribution to the pion form factor from light-cone
QCD sum rules
            \\ }
\vskip1cm
 {\Large Vladimir~Braun} $\footnote { On leave of absence from
St.Petersburg Nuclear
Physics Institute, 188350 Gatchina, Russia}$ \\
\vskip0.2cm
       Max-Planck-Institut f\"ur Physik   \\
       -- Werner-Heisenberg-Institut -- \\
        D--80805, Munich (Fed. Rep. Germany)\\
and\\
\vskip0.5cm
 {\Large Igor~Halperin}
 \\
\vskip0.2cm
       School of Physics and Astronomy   \\
       Raymond and Beverly Sackler Faculty of Exact Sciences \\
       Tel-Aviv University, Tel-Aviv 69978, Israel \\

\vskip1cm
{\Large Abstract:\\}
\parbox[t]{\textwidth}{
We propose a simple method to calculate the pion form factor at
not very large momentum transfers, which combines the technique
of the QCD sum rules with the description of the pion in terms
of the set of wave functions of increasing twist. This approach
allows one to calculate the soft (end point) contribution to the
form factor in a largely model-independent way.
Our results confirm existing expectations that the soft contribution
remains important at least up to the momentum transfers of order
10 GeV$^2$, and suggest that it comes from the
region of relatively small transverse separations of order
1 GeV$^{-1}$.
}
\\ \vspace{1.0cm}
{\em submitted to Physics Letters B }
\end{center}
                                                \end{titlepage}

\newpage
{\bf\large 1.}\hspace{0.5cm}
Hard exclusive processes in QCD  \cite{BLreport}
are attracting continuous interest for already two decades.
In difference to inclusive reactions, like the deep inelastic
scattering, exclusive processes are selective to the partonic
content of the participating hadron. Apriory, one could consider two
different possibilities to transfer a large momentum to a hadron.
One is the so-called Feynman mechanism, in which large momentum
transfer selects the configuration, in which one parton carries
almost all the momentum of the hadron. The transverse size of the
remaining ``soft" cloud remains in this case arbitrary. The second
is the hard rescattering mechanism, in which large momentum
transfer selects configurations with a small transverse size and
a minimal number of Fock constituents. In this case the momentum
fraction  carried by the interacting parton (quark) remains
an average one. The hard rescattering mechanism involves a hard
gluon exchange,
and can be written in the factorized form \cite{exclusive}.

It has been proved \cite{exclusive} that the hard rescattering
mechanism is the leading one at asymptotically large $Q^2$, and
yields the pion form factor
\beq
  F_\pi(Q^2) = \frac{8}{9} \pi \alpha_s\frac{f_\pi^2}{Q^2}
 \left|\int_0^1\frac{du}{1-u} \phi_\pi(u)\right|^2 \,.
\label{Fasym}
\eeq
Here $\phi_\pi(u)$ is the pion wave function of the leading twist,
which describes the distribution of the valence pion constituents
in the longitudinal momentum. Note that convergence of the
integral in (\ref{Fasym}) requires that the pion wave function
decreases at $u\rightarrow 1$, and the crucial point in establishing
the formula in (\ref{Fasym}) was the proof \cite{exclusive}
that at asymptotically
large $Q^2$ the wave function is given by the simple formula
\beq
   \phi^{(as)}_\pi (u) = 6 u (1-u)\,.
\label{phias}
\eeq
The results in (\ref{Fasym}),(\ref{phias}) belong to the most
important and most rigorously proved statements in QCD.

At large but finite momentum transfers there might be a number
of corrections to the hard rescattering formula in (\ref{Fasym}),
and till now  there was a much more
moderate progress in understanding whether available energies could
be treated as asymptotic ones.
 An attempt to describe the data for the pion form
factor starting from $ Q^{2} \geq 3$  GeV$^{2} $ by the contribution of
hard rescattering alone,
implies that the low energy pion wave function
 must be very different
from its asymptotic form, an issue which has been put to fore and
studied in detail by Chernyak and Zhitnitsky \cite{CZreport}.
Using the QCD sum rule approach they have shown that the pion
wave function at a low scale is wider than the asymptotic one,
and proposed a model
\beq
\phi_\pi^{(CZ)}(u,\mu\sim 500\,\mbox{MeV}) = 30 u(1-u)(2u-1)^2\,,
\label{CZWF}
\eeq
which has a peculiar ``humped" profile, with a zero in the middle
point, corresponding to the symmetric configuration where the
quark and the antiquark carry equal momenta.
However, there exists a number of
arguments that force to doubt the assertion about the dominance
of the hard-scattering mechanism in the region of available
$Q^2\sim 1-10$ GeV$^2$.
 The well known point
of Isgur and Llewellyn-Smith \cite{ILS} is that the wave functions
of the type suggested in \cite{CZreport} strongly emphasize the
contribution of the end-point
region of large $u$ in (\ref{Fasym}), where
the virtuality of the gluon is not enough to justify the perturbative
treatment.
This contribution of large $u$, $1-u\sim m^2/Q^2$ (where $m^2$ is a
certain hadronic scale) corresponds to the Feynman mechanism to
transfer the large momentum, and should be treated separately.
At present, there is an increasing evidence that this
contribution is numerically important up to very high $Q^2$, although
it is down by an extra power $1/Q^2$ in the asymptotics.

Using the QCD sum rule approach \cite{SVZ} it has been shown
\cite{IS,NR}
that the pion form factor at $Q^2\sim 1-2$ GeV$^2$ is practically
saturated by the Feynman-type contribution.  Unfortunately, this
method in its standard form cannot be applied for higher values of
$Q^2$, since it involves the expansion in the contributions of
vacuum condensates, which coefficients appear to be enhanced
by increasing powers of $Q^2$. Thus, at sufficiently large
momentum transfers the expansion breaks down.
In attempt to cure this problem Radyushkin and collaborators
\cite{MR,Rad}
have suggested to resum the series of power corrections
by introducing the nonlocal extension of the concept of the vacuum
condensates, which takes into account the final correlation length
in the QCD vacuum. Results of \cite{Rad} indicate the dominating
role of the soft contribution at least up to $Q^2 \sim 10$ GeV$^2$.

 The approach of \cite{Rad} clearly demonstrates the origin
of difficulties in the standard QCD sum rule calculations, but may
receive objections concerning its theoretical accuracy,
 since not
 all high-order contributions can consistently be taken into account
in this way, and also the parametrization of the nonlocal condensates
is essentially model-dependent.
In this letter we suggest an alternative approach to the calculation
of the pion form factor at not very large values of $Q^2$, which
seems to be free from both the difficulties of the standard
QCD sum rule method, and the ambiguities involved in its
extension in \cite{MR,Rad}.
Our method essentially follows earlier works \cite{CS}--\cite{BKR},
where the QCD sum rule approach has been modified to incorporate
the operator product expansion in powers of the deviation from the
light-cone (in contrast to the short distance expansion in \cite{SVZ}).

\bigskip
{\bf\large 2.}\hspace{0.5cm}
The idea is to combine
the standard technique  for the study of
hard exclusive processes and the QCD sum rule method.
To this purpose, we consider the correlation function
\beq
T_{\mu \nu}(p,q)\;=\;i \int dx \exp (iqx) \langle 0 | T \{ j_{\mu}^{5}(0)
j_{\nu}^{em}(x) \} | \pi^{+}(p) \rangle  \;,
\label{CF}
\eeq
where $ j^{5}_{\mu} = \bar{d} \gmmu \gmf u $ and $ j^{em}_{\nu}=
e_{u} \bar{u} \gmnu u + e_{d} \bar{d} \gmnu d $
is the electromagnetic current.
 At large Euclidian momenta $ (p-q)^{2} $ and $ q^{2} $
this correlation function can be calculated in QCD, in the precise
analogy with the  calculation of the $\pi\gamma^*\gamma^*$ form factor
(for different virtualities of photons), \cite{CZreport}.
The leading contribution is written in terms of the pion wave function
of the leading twist
\beq
 T_{\mu \nu}(p,q)\;=\; 2i f_\pi p_\mu p_\nu
\int_0^1 du \frac{u\phi_\pi(u)}{(1-u)Q^2- u(q-p)^2} +\ldots\,,
\label{a}
\eeq
where $f_\pi$ is the pion decay constant, $Q^2=-q^2$,
and the ellipses stand
for the contributions of other Lorentz structures.
On the other hand,
the dispersion relation over $ (p-q)^{2} $ relates the correlation
function (\ref{CF}) to the pion form factor:
\beq
T_{ \mu \nu} = i f_{ \pi} (p-q)_{ \mu} \frac{1}{ m^{2}_{ \pi} -
(p-q)^{2} } 2 F_{ \pi}(q^{2})p_{ \nu} + \ldots\,,
\label{b}
\eeq
where the dots stand for higher resonances  and the continuum
contributions.
Matching between the representations in (\ref{a}) and  (\ref{b})
at no-so-large Euclidian $-(q-p)^2\sim 1$ GeV$^2$,
we obtain a sum rule for the pion form factor.

To this end, we use the standard concept of
duality,
which tells that the pion occupies the ``region of duality"
in the invariant mass of the $\bar q q $ pair, up to a certain
threshold $s_0\sim 0.7-0.8$ GeV$^2$. Note that the formula in
(\ref{a}) can be rewritten as the dispersion relation in $(p-q)^2$,
with $s=(1-u)Q^2/u$
being the mass of the intermediate state.
To pick up the contribution of the pion, we cut the dispersion integral
at $s=s_0$, which translates to a low bound for the integral over $u$:
$u_{min}= Q^2/(s_0+Q^2)$.
Following \cite{SVZ}, we also use the Borel transformation to
convert the power suppression of higher mass contributions in
the dispersion integral to the exponential suppression
\bea
\frac{u}{(1-u)Q^2-u(q-p)^2}&\rightarrow &
            \exp\left\{-\frac{(1-u)Q^2}{uM^2}\right\}\,,
\nonumber\\
\frac{1}{m_\pi^2-(q-p)^2}&\rightarrow &
            \exp\left\{-\frac{m_\pi^2}{M^2}\right\}\,,
\label{Borel}
\eea
where $M^2$ is the new variable (the Borel parameter).  In what follows
we put the pion mass to zero.

Thus, the sum rule arises
\beq
F_{\pi}(Q^{2})   =
\int_{0}^{1} du \,\phi_\pi(u)
\exp\left\{-\frac{(1-u)Q^2}{uM^2}\right\}
\Theta\left(  u - \frac{Q^{2}}{s_{0}+Q^{2}}\right)\,,
\label{SR1}
\eeq
which should be satisfied at the values of the Borel parameter $M^2$
of order 1 GeV$^2$.  The pion wave function in (\ref{SR1}) should
be taken at a low normalization scale, of order of the Borel
parameter.

In what follows we shall complement the sum rules in (\ref{SR1})
by contributions of higher twist. Before doing this,
and before going over to the numerical analysis, let us
study the behavior of the sum rule in the limit of large
momentum transfers $Q^2\rightarrow\infty$.

Because of the $\Theta$-function, the integration region in
(\ref{SR1})  is restricted to values $1-u < s_0/(s_0+Q^2) \rightarrow 0$.
Thus, the form factor is sensitive to the wave function
in the highly asymmetrical configuration, where the scattered quark
carries almost all the pion momentum. According to the general analysis
in \cite{exclusive,CZreport} the behavior of the pion wave function
in this region coincides with the asymptotic behavior in (\ref{phias}),
$\phi_\pi(u) \stackrel{u\rightarrow 1}{\sim} 1-u$.
Thus, asymptotically, the sum rule in (\ref{SR1}) yields
\beq
 F_{\pi}(Q^{2}) \sim \frac{\phi'_\pi(0)}{Q^4}
 \int_0^{s_0} s^2 ds\, e^{-s/M^2}\,,
\label{Fsoft1}
\eeq
where $\phi'_\pi(0)=(d/du)\phi_\pi(u)|_{u=0}$. (We have made use of the
symmetry $\phi_\pi(u)=\phi_\pi(1-u)$).

For the asymptotic wave function (\ref{phias}) $\phi'_\pi(0) =6$.
For the model suggested by Chernyak and Zhitnitsky  (\ref{CZWF})
$\phi_\pi^{'(CZ)}(0) = 30$, and thus it
gives rise to a much bigger soft contribution to the form factor.
This is in agreement with the observation by Isgur and Llewellyn-Smith
\cite{ILS} that ``realistic"  model wave functions, which allow for
the description of the data on pion form factor at virtualities of
order a few GeV by the hard rescattering mechanism alone, in fact
greatly enhance the soft contribution coming from the end point
region.

\bigskip
{\bf\large 3.}\hspace{0.5cm}
Now we are in a position to take into account the corrections
to the sum rule in (\ref{SR1}), induced by the pion wave functions
of the nonleading twist 4. These corrections correspond to the
contributions to the correlation function in (\ref{CF}), which
are suppressed by an extra power of the large momentum $(p-q)^2$ or
$q^2$. Physically, these corrections have two sources. One of them
is the transverse momentum of quarks in the leading order diagram
in Fig. 2a. The second source are contributions of higher Fock
states in the pion wave function --- containing gluon fields in
addition to the quark and antiquark ones.  In fact, these two effects
are indistinguishable thanks to the equations of motion,
which allow one to eliminate all transverse degrees of freedom at the
cost of introducing the higher Fock components.
In this letter we cannot give all the details of the calculation,
which becomes rather technical. A suitable  formalism for the
operator product expansion of the correlation functions beyond
the leading twist has been developed in \cite{BB88}.

The complete contribution of the diagram in Fig. 1a to the twist-4
accuracy reads
\bea
\Pi_{(2a)} &=& 2 i p_\mu p_\nu
f_{ \pi} \int_{0}^{1} du \Bigg\{
- \frac{ u \phipi }{ (q-up)^{2}} - 4u\,\frac{g_{1}(u) + G_{2}(u)}
{(q-up)^{4}}
+ 2 u^{2}  \frac{g_{2}(u)}{(q-up)^{4}}
  \Bigg     \}
+\ldots \,,
\label{diag2a}
\eea
where we have introduced the pion wave functions of
twist 2 and 4 defined by the matrix element \cite{BF2}
\bea
\lo \bar{d}(0) \gmmu \gmf u(x) | \pi(p) \rangle & = & if_{\pi}p_{\mu}
\int_{0}^{1}
du\, e^{-iupx} ( \phipi  +  x^{2}g_{1}(u) + O(x^{4}))  \nonumber    \\
      &&\mbox{}+ f_{\pi}(x_{\mu}-
\frac{x^{2}p_{ \mu}}{px}) \int_{0}^{1} du\, e^{-iupx} g_{2}(u) +
\ldots \,,
\label{WF2}
\eea
and
all logarithmic dependence on $ x^{2}$ is implicitly included
in the wave functions. The function $ G_{2} $ in (\ref{diag2a})
is defined as
$ g_{2}(u) = -(d/du)\, G_{2}(u) $.
The diagram shown in Fig. 1b produces an additional contribution
 ($ \alpha_{3}= 1- \alpha_{1}- \alpha_{2} $)
\bea
\Pi_{(2b)} &=& 2i f_{ \pi} \int_{0}^{1} \frac{u\,du }{ (q-up)^{4}}
\int_{0}^{u}
d \alpha_{1} \int_{0}^{ \bar{u}} d \alpha_{2} \Bigg
\{
\frac{ \Psi_{ \parallel} + 2 \Psi_{ \perp}}{ \alpha_{3}} \nonumber \\
&&\mbox{}+ \frac{1-2u +\alpha_{1}- \alpha_{2}}{
\alpha_{3}^{2}}
( \Phi_{ \parallel} + 2 \Phi_{ \perp} ) \Bigg\}\,,
\label{diag2b}
\eea
where we have followed \cite{BF2} in the definition of
three-particle wave functions of twist 4:
\bea
\lo \bar{d}(-x) \gmmu \gmf g G_{ \alpha \beta }(vx) u(x) |
\pi(p) \rangle   =   p_{ \mu} ( p_{ \alpha} x_{ \beta} - p_{ \beta}
x_{ \alpha} ) \frac{1}{px} f_{ \pi} \int D \alpha \Phi_{ \parallel}
( \alpha_{i}) e^{ -ipx( \alpha_{1} - \alpha_{2} + v \alpha_{3}) }
   \nonumber  \\
+\left[p_{ \beta}( \delta_{ \alpha \mu} -
\frac{ x_{ \alpha} p_{ \mu}}{ px} ) - p_{ \alpha} ( \delta_{ \beta
\mu } - \frac{ x_{ \beta} p_{ \mu}}{ px} ) \right] f_{ \pi}
\int D \alpha \Phi_{ \perp} ( \alpha_{i}) e^{ -ipx( \alpha_{1}-
\alpha_{2} + v \alpha_{3})}\,.
\label{WF3}
\eea
The wave functions
 $ \Psi_{ \parallel}$ and $\Psi_{ \perp} $ are defined similar to
(\ref{WF3}),
with the substitution $ ( \gmf g G_{ \alpha \beta} ) \rightarrow
( i g \tilde G_{ \alpha \beta} ) $.

A systematic study of the higher twist wave
functions has been done in the work \cite{BF2}, and makes use of
the expansion in representations of the collinear
conformal group SO(2,1), which is a subgroup of full conformal
group acting on the light-cone. The asymptotic wave functions are
defined as contributions of operators with the lowest conformal
spin. The set of wave functions suggested in \cite{BF2} includes
also the corrections corresponding to the operators with the
next-to-leading conformal spin, which numerical values are
calculated by the QCD sum rule method. The results for twist 4
wave functions are
(hereafter $ \bar{u} = 1-u $):
\bea
\Phi _{\parallel}(\alpha _{i}) &=& 120 \varepsilon \delta ^{2}
(\alpha_{1}- \alpha_{2}) \alpha_{1} \alpha_{2} \alpha_{3}\,,
 \nonumber \\
\Psi _{\parallel}(\alpha_{i}) &=& -120 \delta ^{2}
\alpha_{1}\alpha_{2}\alpha_{3}
\Big[ \frac{1}{3} + \varepsilon (1- 3 \alpha_3 ) \Big]\,,
\nonumber \\
\Phi_{\perp}(\alpha_{i}) &=& 30 \delta^{2} (\alpha_{1} - \alpha_{2})
\alpha_{3}^{2}
\Big[ \frac{1}{3} + 2 \varepsilon ( 1- 2 \alpha_{3})\Big]\,,\\
\Psi_{\perp}( \alpha_{i}) &=& 30 \delta^{2} \alpha_{3}^{2} (1- \alpha_{3}
)\Big[ \frac{1}{3} + 2 \varepsilon ( 1- 2 \alpha_{3})\Big]\,,
 \nonumber \\
g_{1}(u) &=& \frac{25}{6} \delta^2 \bar u^2 u^2 +
\varepsilon \delta^2\Big[ \bar u u (2+13 \bar u u)
\nonumber\\&&\mbox{}+10 u^3  (2-3 u+\frac{6}{5}u^2) \ln u
+10\bar u^3 (2-3\bar u+\frac{6}{5}\bar u^2)\ln\bar u\Big]\,,
 \nonumber  \\
g_{2}(u)&=& \frac{10}{3} \delta ^{2}\bar{u} u ( u- \bar{u} ) \,,
\nonumber \\
G_2(u)&=& \frac{5}{3}\delta^2 u^2 \bar u^2 \,,
\nonumber \\
\delta ^{2} &\simeq &0.2\;\mbox{\rm GeV}^2
 \;,\; \varepsilon \simeq 0.5 \nonumber\,.
\label{WF4}
\eea
Adding the higher-twist
contributions in (\ref{diag2a}) and (\ref{diag2b}) to
the sum rule in (\ref{SR1}) we arrive at
\bea
\lefteqn{F_{\pi}(Q^{2})  =
\int_{0}^{1} du \,
\exp\left[-\frac{\bar u Q^2}{uM^2}\right]
\Bigg\{\, \phipi  - \frac{4}{uM^{2}}(g_{1}(u) + G_{2}(u))
               + \frac{2}{M^{2}}g_{2}(u)  }
      \\ &&\mbox{}
           + \frac{1}{uM^{2}} \int_{0}^{u}
\!d \alpha_{1}\! \int_{0}^{ \bar{u}}\! d \alpha_{2} \Bigg[
\frac{ \Psi_{ \parallel} + 2 \Psi_{ \perp}}{ \alpha_{3}}
+ \frac{1-2u +\alpha_{1}- \alpha_{2}}{
\alpha_{3}^{2}}
( \Phi_{ \parallel} + 2 \Phi_{ \perp} ) \Bigg]\Bigg\}
\Theta\left(  u - \frac{Q^{2}}{s_{0}+Q^{2}}\right)\,,
  \nonumber
\label{SR}
\eea
which presents our final result.
Note that the higher-twist contributions
are suppressed by a power of the Borel parameter $M^2$, as expected.

With the particular expressions (\ref{WF4}) the integrals of the
three-particle wave functions can be taken analytically, yielding
\bea
\int_{0}^{u}d \alpha_{1} \int_{0}^{ \bar{u}} d \alpha_{2}
\frac{ \Psi_{ \parallel} + 2 \Psi_{ \perp}}{ \alpha_{3}} &=&
\frac{10}{3}\delta^2 \bar u u (1-2 \bar u u)\,,
\nonumber\\
\int_{0}^{u}d \alpha_{1} \int_{0}^{ \bar{u}} d \alpha_{2}
\frac{1-2u +\alpha_{1}- \alpha_{2}}{\alpha_{3}^{2}}
( \Phi_{ \parallel} + 2 \Phi_{ \perp} ) &=&
  -2 g_1(u) -\frac{10}{3}\delta^2 \bar u u
\Big( 1-\frac{15}{2}\bar u u \Big)\,.
\label{alpha}
\eea
Note that the term $\sim \bar u u$ in (\ref{alpha}) cancel in the
sum rule (\ref{SR}), and thus contributions of three-particle wave
functions turn out to be of order $1/Q^6$, i.e. are suppressed by an
additional power of $1/Q^2$. However, the twist 4 contributions still
 survive in the high-$Q^2$ limit
due to the contribution of $g_2$, which
 yields the same asymptotic
behavior $\sim 1/Q^4$ as the leading twist contribution in
(\ref{Fsoft1}).

\bigskip
{\bf\large 4.}\hspace{0.5cm}
We turn now to the numerical analysis. Apart from the wave functions,
the sum rule in (\ref{SR}) depends on the value of the continuum
threshold $s_0$, and on the Borel parameter $M^2$. We take
$s_0=0.7-0.8$ GeV$^2$ \cite{SVZ} and vary $M^2$ in the interval
1 -- 2 GeV$^2$, which is the expected stability region.
The results are shown in Fig.2. In Fig. 2a  we plot the value
of $Q^2 F_\pi(Q^2)$ as a function of $Q^2$ for $s_0=0.7$ GeV$^2$ and
$s_0=0.8$ GeV$^2$ and for different choices of
the leading twist pion wave function
$\phi_\pi(u)$: asymptotic wave function (\ref{phias})
and the Chernyak-Zhitnitsky
model (\ref{CZWF}). Since this model in fact
refers to a substantially lower normalization point than the typical
value of the Borel parameter in the sum rule, we give the results also
for the Chernyak-Zhitnitsky wave function rescaled to
 $\mu^2 \sim 1-2$ GeV$^2$
\beq
\phi_\pi^{(CZ)}(u,\mu\sim 1 \mbox{GeV}) = 6 u(1-u)
\Big[1+0.44 C^{3/2}_2(2u-1)\Big]\,.
\label{CZWF1}
\eeq
The wave function in (\ref{CZWF1}) corresponds to the value of the
second moment $\langle (2u-1)^2\rangle = 0.35$ \cite{BF1}, which is to
be
compared to $\langle (2u-1)^2\rangle = 0.43$ for (\ref{CZWF}).
We remind that for the asymptotical wave function
$\langle (2u-1)^2\rangle = 0.2$.
 The contribution of wave functions of twist 4
does not exceed 20\%, and these wave functions are not far from
their asymptotic form. Thus possible inaccuracy in the model
wave functions in (\ref{WF4}) does not have any noticeable effect.
The stability of the sum rule (\ref{SR}) to the choice of the
Borel parameter is illustrated in Fig. 2b for several values of
$Q^2$.

It is seen that the soft contribution to the pion form factor
clearly dominates at $Q^2\sim 1-3 $ GeV$^2$ and constitutes about
15--30\%
of the experimental value at $Q^2 \simeq 10 $ GeV$^2$
(for the asymptotic wave function).
For the Chernyak--Zhitnitsky model, the soft contribution increases
substantially.

Within our approach, the hard gluon exchange contribution
originates from the radiative correction to the contribution of
the leading twist, see diagram in Fig. 1c.
This contribution is not restricted to the
end-point region, and thus has no power-like $1/Q^2$ suppression.
Its explicit calculation
goes beyond the tasks of this letter. As a rough estimate, one can
use the expression in (\ref{Fasym}), yielding
$Q^2 F_\pi(Q^2)_{\mbox{\scriptsize hard}} \simeq 0.15  $ and
$Q^2 F_\pi(Q^2)_{\mbox{\scriptsize hard}} \simeq 0.3$ for the asymptotic
wave function and the Chernyak--Zhitnitsky model, respectively.
One sees that the full answer for the pion form factor, given by
the sum of the soft and hard contributions, is likely
to overshoot the data, if one uses the Chernyak--Zhitnitsky model.

Main lesson to be learnt from our calculation is that the
soft contribution to the pion form factor decreases very slowly
with $Q^2$ and is important in the whole region of momentum
transfers, which are available at present. This conclusion
is in full agreement with the results of \cite{ILS,Rad}, although
our argumentation is different.

\bigskip
{\bf\large 5.}\hspace{0.5cm}
The method described above is quite general, and can be applied to
different form factors as well. As a one more example,
we calculate here the soft contribution to the transition
form factor $ \gamma \pi \rho $.
For the $ | \lambda | =1 \; \; \rho $ - meson ($ \rhop $ hereafter)
the transition form factor is defined as
\beq
\langle \rhop (p_{1})| j_{ \mu}^{em} | \pi (p_2) \rangle
= \varep p_{1}^{\lambda} p_{2}^{ \sigma} \varepsilon _{ \perp}^{ \rho}
F_{ \pi \rho }( Q^{2})    \; ,
\label{rhopigamma}
\eeq
where $Q^{2} = - (p_{1} -p_{2} )^{2} $ and $ \varepsilon _{ \perp}
^{ \rho} $ stands for the polarization vector of the $ \rho $. This
process is clearly due to non-leading twist effects \cite{CZreport}
as long as
it is related with the helicity flipping. As a result, the hard
rescattering diagram yield the asymptotic behavior $ \frho \sim
1/Q^{4} $ \cite{CZreport}.
As we will show, the soft contribution
exhibits  the same asymptotic dependence
$ 1/Q^{4} $, and is of the same order, therefore, as the hard
contribution
 (up to the Sudakov suppression, which is unlikely to be
 important at moderate values of $Q^2$).

We consider the correlation function
\bea
A(p,q) &=& i \int dx e^{iqx} \lo T \{ \bar{d}(0) \sigrh
u(0) j_{ \mu}^{em}(x) \} | \pi (p) \rangle    \nonumber  \\
       &=& ( \varep q_{ \xi} - \varepsilon_{ \mu \xi \lambda
\sigma } q_{ \nu} ) q_{ \lambda} p_{ \sigma} \Pi ( q^{2}, (p-q)^{2}) +...
\label{CFrho}
\eea
which contains a contribution of interest of the $\rhop$-meson
\beq
\Pi ( q^{2},(p-q)^{2}) = f_{ \rho}^{T} \frac{ \frho }{ m_{ \rho}^{2}
- (p-q)^{2}} \; .
\eeq
Here  $ f_{ \rho}^{T} $ is the $ \rhop $-meson decay constant:
\beq
\lo \bar{d}(0) \sigrh \gmf u(x) | \pi (p) \rangle
= i( \varepsilon _{ \xi}^{ \perp} p_{ \nu} - \varepsilon_{ \nu}
^{ \perp} p_{ \xi} ) f_{ \rho}^{T}\,.
\eeq
Calculation of the diagram in Fig.1a yields in this case
\beq
\Pi ( q^{2}, (p-q)^{2}) = - (e_{u}+ e_{d}) \frac{ f_{\pi} m_{ \pi}^{2}}
{3 ( m_{u} + m_{d}) } \int_{0}^{1} du \frac{ \varphi_{ \sigma}(u)}
{ (q-up)^{4}}  \; ,
\eeq
where $ \varphi_{ \sigma}(u) $ is the pion wave function
 of twist 3 \cite{BF2}
\beq
\lo \bar{d}(0) \sigrh \gmf u(x) | \pi (p) \rangle
= \frac{ i f_{ \pi} m_{\pi}^{2}}{ 6 ( m_{u}+ m_{d} )}
( p_{ \xi} x_{ \nu} - p_{ \nu} x_{ \xi} )
\int_{0}^{1} du e^{ -iupx} \varphi_{ \sigma} (u)\,.
\label{WFsigma}
\eeq
It has been shown in \cite{BF2} that the wave function
$ \varphi_{ \sigma}(u) $ is close to its asymptotic form
$ \varphi_{ \sigma}(u) =6 u (1-u)$.

It is easy to check that the diagram of Fig.1b does not contribute to
the Lorentz
structure of interest. Then, to the twist 3 accuracy, we arrive at
the very simple sum rule
\beq
\frho = ( e_{u} + e_{d} ) \frac{ 2 \langle \bar{ \psi} \psi \rangle }
{3 f_{\pi}
f_{\rho}^{T}} \frac{ e^{ m_{\rho}^{2} / M^{2} } } {M^{2}}
\int_{0}^{1} du \,e^{ - \frac{ \bar{u} Q^{2}}{u M^{2}}}
\frac{ \varsig }{ u^{2}} \Theta ( u - \frac{ Q^{2}}{ s_{0} + Q^{2}} )
\,,
\label{SRrho}
\eeq
in which we have replaced the factor appearing in the normalization
of the wave function $\phi_\sigma$ (\ref{WFsigma}) by the quark
condensate $\langle\bar q q\rangle \simeq -(250$~MeV$)^3$.
Following \cite{SVZ,CZreport} we use the values $ s_{0}=1.5$ GeV$^2$
and  $ f_{\rho}^{T} \simeq 200$ MeV
for the
continuum threshold in the $\rho$-meson channel, and
the $\rho$-meson coupling, respectively.
The results are shown in Fig.~3. Using them we obtain an estimate
for the $ \Psi \rightarrow \gamma \rightarrow \pi^{0} \omega $
decay rate
($ F_{\pi \omega}(Q^{2}) = 3 \frho $ due to the isospin symmetry)
\beq
Br \; (  \Psi \rightarrow \gamma \rightarrow \pi^{0} \omega \;
/ \Psi \rightarrow  e^{+} e^{-} ) = \frac{9}{32} ( M_{\Psi} F_{\pi
\omega} ( M_{\Psi}^{2} ))^{2} \simeq (4\pm 2) \times 10^{-4}\,.
\label{Brpsipiom}
\eeq
A relatively large error is due to a poor stability of the sum
rule in this case. The number in (\ref{Brpsipiom})
 appears to be in good agreement to the
 experimental number $(4.2\pm 0.6)\times 10^{-4}$ \cite{PDT}.

The contribution to this form factor of the hard rescattering  has been
 calculated by Chernyak and
Zhitnitsky \cite{CZreport}, using the leading twist pion wave function
in (\ref{CZWF}), and three-particle wave functions of the $\rho$-meson
of  nonleading twist.
It has the same functional dependence $\sim 1/Q^4$,
and approximately the same numerical value as the soft contribution
which we have calculated here.
A simple patching them together
would yield the branching ratio
$ \Psi \rightarrow \pi^{0} \omega $
several times above the data. To our opinion, the  hard contribution
to this decay given in \cite{CZreport} is
strongly overestimated.

\bigskip
{\bf\large 6.}\hspace{0.5cm}
In this letter we have suggested a simple method to calculate
the pion form factor in the region of intermediate momentum
transfers, which is essentially a hybrid of the standard
QCD sum rule approach and the conventional expansion in
terms of the pion wave functions. Its value is in the possibility
to estimate the soft (end point) contribution to the form factor
in a model independent way, which is a problem of acute interest.
The main advantage compared to the standard QCD sum rule
calculation \cite{IS,NR}  is that the ``light-cone sum rules"
suggested in this paper remain well-defined in the limit
$Q^2\rightarrow\infty$, and is related to the fact that the
parameter of the expansion in our sum rules is the twist
of relevant operators, but not their dimension as in the standard
sum rules.  In this way contributions of various local operators
are resummed in the set of wave functions of increasing twist,
the end-point behavior of which is known from general arguments.
In effect, explicit factors $\sim Q^2$ which may appear in the
calculation of higher twist contributions will be compensated
by factors $\sim1/Q^2$ originating from a more fast decrease
of higher twist wave functions at $u\rightarrow 1$ compared to
the leading twist ones.
The physical reason for disappearance of the contributions enhanced
by powers of $Q^2$ is in our approach the same as in the calculation
involving the nonlocal condensates in \cite{Rad}. However, our
method is practically model-independent.

Our main result is the calculation of the soft contribution to the
form factor, which turns out to be large at least up to
$Q^2\sim 10$~GeV$^2$. In agreement to \cite{ILS} we find that
this contribution depends strongly on the shape of the pion wave
function. Patching together the contribution of hard rescattering
and the soft contribution, we find that the model by Chernyak and
Zhitnitsky is likely to overshoot the data.
Combining this result with the criticism in \cite{MR,BF1}, we
conclude that there is increasing evidence, coming from
different calculations, that the true low energy pion wave function
is not that much different from its asymptotic form, as proposed
in \cite{CZreport}.

On the evidence of
an impressive calculation of Sudakov-type
double-logarithmic corrections to the contribution of the hard
rescattering,
it has been argued in \cite{LS} that the end-point contribution
to the pion form factor
is strongly suppressed already at moderate $Q^2\sim 10$~GeV$^2$.
Radiative corrections to the correlation function in (\ref{CF})
can involve logarithms of the type $\ln(q^2/(q-p)^2)$, but not
$\ln(Q^2/\Lambda_{QCD}^2)$, since it is IR-protected.
Hence the corrections to the sum rule can only be accompanied
by logs like $\ln(Q^2/s_0), \ln(Q^2/M^2)$ which never become
large (at moderate momentum transfers). Thus, the Sudakov exponential
suppression is not likely to occur in our sum rules.
To our opinion, the significance of Sudakov corrections is
 overestimated in \cite{LS}.  The reason is that
the effective transverse momentum, generated by the Sudakov suppression,
should be compared not to the Compton wave length of the pion,
of order 1/200 MeV, but to the average transverse separation
of the quark and antiquark in the particular configuration which
dominates the
soft contribution to the form factor.
Note in this respect, that the quark-antiquark separation in a
``free" pion does not have any physical meaning beyond the leading
twist accuracy, since effects of transverse degrees of freedom
can be rewritten in terms of higher Fock components in the wave
function thanks to the equations of motion.
For the particular hard process, however, the question of relevant
transverse distances is well-defined.
In the case of the correlation function in (\ref{CF}),
 the characteristic transverse separation
between the quark and the antiquark is given by the deviation from
the light-cone  $x^2\sim (q-up)^2$, as can easily be checked by
an explicit calculation in light-cone coordinates in the position
space. After the Borel transformation, $(q-up)^2$  goes into $uM^2$,
so that the characteristic transverse separations yielding
the form factor in Fig.~2 are of the order
\beq
   x_\perp^2 \sim \frac{1}{uM^2}\,.
\label{size}
\eeq
Note that $1/(uM^2)$  is exactly the expansion parameter in
our calculation, which controls the size of higher-twist
corrections, and in the working region of the sum rule is
of  order $1/s_0 \sim (0.2-0.3$~fm$)^2$.
To our opinion, it is this scale rather than
$\Lambda_{QCD}$
which should serve as the IR cutoff in the calculation in \cite{LS}.
Since the average value of $u$ under the integral in the sum rule
increases slightly with $Q^2$, one may speculate that the
relevant transverse size and the importance of higher twist
contributions are slightly decreasing.

Thus, the soft contribution to the pion form factor
comes from configurations with a much smaller transverse size
than the poin electromagnetic radius $\sim 0.65 $ fm,
which is dominated by contributions of multiparton
states.
At distances $\sim 0.2-0.3$~fm  the strong coupling is already not
large, and the application of perturbation theory to the calculation
of the contribution of the hard gluon exchange can be justified.
However, this contribution must be complemented by the contribution
of Feynman type, coming from the end-point region.  Our conclusions
essentially support the picture described in \cite{Rhistory}.

\bigskip
We gratefully acknowledge stimulating discussions with
L.L.~Frankfurt and \hfill\break
A.V.~Radyushkin on the number of subjects related
to this study.
Our special thanks are due to V.M.~Belyaev for pointing out an error
in the preliminary version of this paper.
\clearpage



\section*{Captions}
\begin{description}
    \item [Fig. 1] Leading contributions to the expansion of the
                   correlation function in (\ref{CF}) in powers
                   of the deviation from the light-cone.
    \item [Fig. 2] The soft (end point) contribution to the pion
                   electromagnetic form factor for the asymptotic
                  pion wave function and for the Chernyak--Zhitnitsky
                  model as a function of $Q^2$ (a) and in dependence
                  on the Borel parameter $M^2$ in the sum rule (b).
                  The solid and dashed curves in Fig.~2a correspond
                  to the calculation with $s_0=0.8$ and $s_0=0.7$
                  GeV$^2$, respectively, and the value of the Borel
                  parameter $M^2=1.5$ GeV$^2$. The curves in Fig.~2b
                  are calculated using  $s_0=0.8$ GeV$^2$ and the
                  asymptotical wave function.
                  Among the pairs of curves marked
                  ``CZ'' the upper ones correspond to the
                  calculation using the Chernyak-Zhitnitsky wave
                 function at the scale $\mu = 500$ MeV,
                 and the  lower ones at $\mu= 1$ GeV,
                 see (\ref{CZWF}) and
                (\ref{CZWF1}), respectively.

    \item [Fig. 3] The transition form factor $\gamma\rho\pi$
                  as a function of the momentum transfer.
\end{description}

\clearpage

\begin{figure}
    \begin{center}
        \begin{picture}(110,140)
        \end{picture}
    \end{center}
    \caption[xxx]{
                   Leading contributions to the expansion of the
                    correlation function in (\ref{CF}) in powers
                    of the deviation from the light-cone.
                    }
   \label{pic.a}
\end{figure}
\begin{figure}
    \begin{center}
          \begin{picture}(120,175)
          \end{picture}
    \end{center}
    \caption[xxx]{
                   The soft (end point) contribution to the pion
                    electromagnetic form factor for the asymptotic
                   pion wave function and for the Chernyak--Zhitnitsky
                   model as a function of $Q^2$ (a) and in dependence
                   on the Borel parameter $M^2$ in the sum rule (b).
                   The solid and dashed curves in Fig.~2a correspond
                   to the calculation with $s_0=0.8$ and $s_0=0.7$
                   GeV$^2$, respectively, and the value of the Borel
                   parameter $M^2=1.5$ GeV$^2$. The curves in Fig.~2b
                   are calculated using  $s_0=0.8$ GeV$^2$
                   and the asymptotical wave function.
                  Among the pairs of curves marked
                  ``CZ'' the upper ones correspond to the
                  calculation using the Chernyak-Zhitnitsky wave
                 function at the scale $\mu = 500$ MeV,
                 and the  lower ones at $\mu= 1$ GeV,
                 see (\ref{CZWF}) and
                (\ref{CZWF1}), respectively.
  }
   \label{pic.b}
\end{figure}
\begin{figure}
    \begin{center}
\begin{picture}(100,105)
\end{picture}
    \end{center}
\caption[xxx]{
                 The transition form factor $\gamma\rho\pi$
                   as a function of the momentum transfer.
 }
   \label{pic.c}
\end{figure}
\clearpage
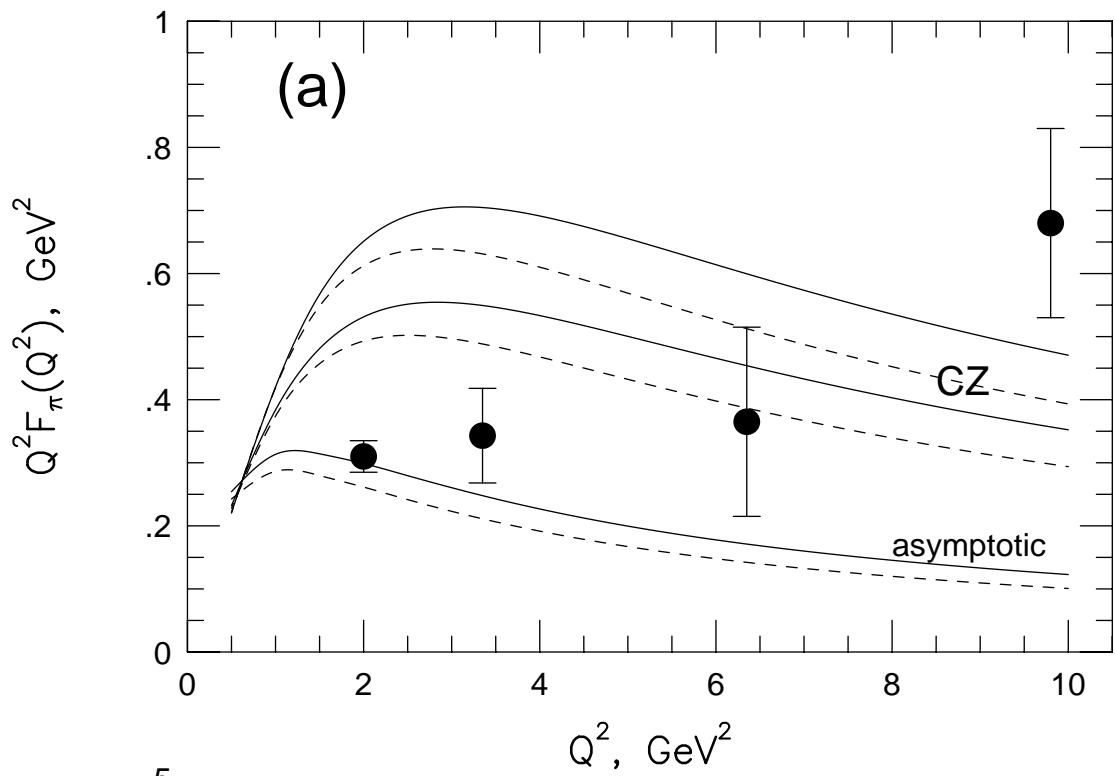
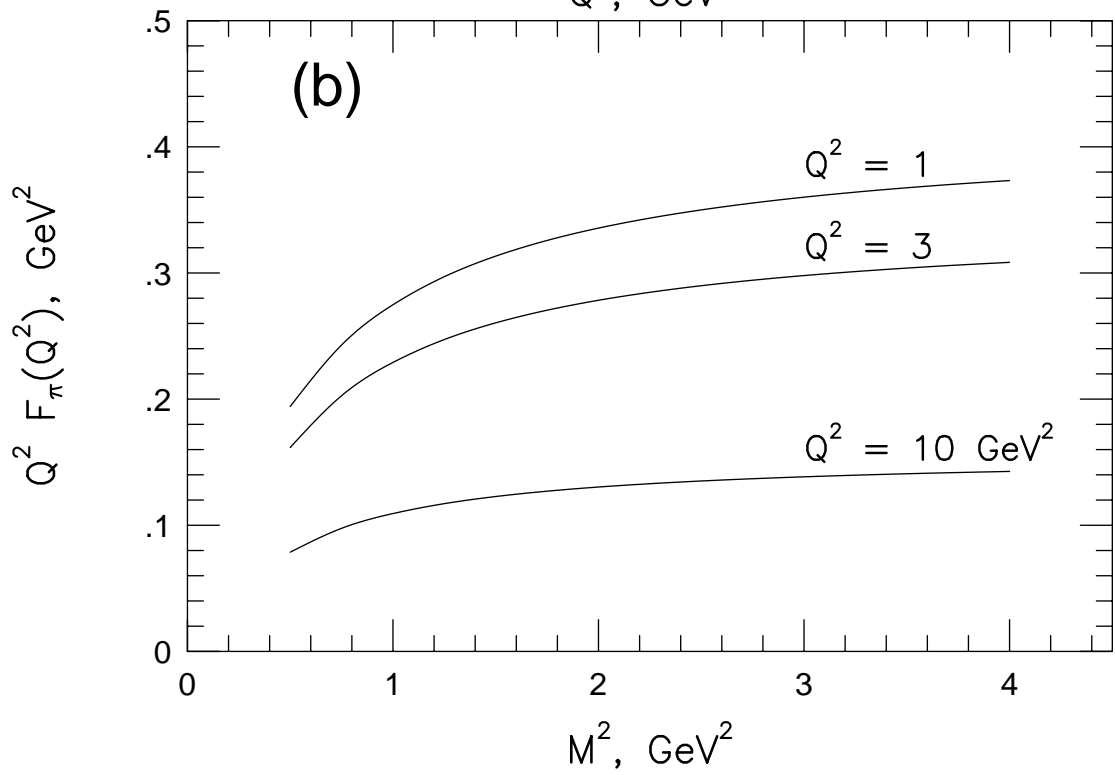
\end{document}